\documentclass[preprint2]{aastex}
\shorttitle{Polarimetry of R CrA Region}
\shortauthors{Choi et al.}

\usepackage{times}
\newcommand{\Jypb}{{Jy beam$^{-1}$}}
\usepackage{epsfig}
\newcommand{\ploteps}[2]{\centerline{\ifdim#2=0mm\epsfig{file=#1}%
                                     \else\epsfig{file=#1,width=#2}\fi}}

\slugcomment{To appear in the Astrophysical Journal}

\begin{document}
\fontsize{10}{10.6}\selectfont
\title{Centimeter Polarimetry of the R Coronae Australis Region}
\author{\sc Minho Choi\altaffilmark{1},
            Ken'ichi Tatematsu\altaffilmark{2},
            Kenji Hamaguchi\altaffilmark{3,4},
            and Jeong-Eun Lee\altaffilmark{5}}
\altaffiltext{1}{International Center for Astrophysics,
                 Korea Astronomy and Space Science Institute,
                 Hwaam 61-1, Yuseong, Daejeon 305-348, South Korea;
                 minho@kasi.re.kr.}
\altaffiltext{2}{National Astronomical Observatory of Japan,
                 2-21-1 Osawa, Mitaka, Tokyo 181-8588, Japan.}
\altaffiltext{3}{Center for Research and Exploration
                 in Space Science and Technology
                 and X-ray Astrophysics Laboratory,
                 NASA Goddard Space Flight Center,
                 Greenbelt, MD 20771.}
\altaffiltext{4}{Department of Physics, University of Maryland,
                 Baltimore County, 1000 Hilltop Circle, Baltimore, MD 21250.}
\altaffiltext{5}{Department of Astronomy and Space Science,
                 Astrophysical Research Center
                 for the Structure and Evolution of the Cosmos,
                 Sejong University, Gunja 98, Gwangjin,
                 Seoul 143-747, South Korea.}
\setcounter{footnote}{5}

\begin{abstract}
\fontsize{10}{10.6}\selectfont
Circularly polarized 3.5 cm continuum emission
was detected toward three radio sources in the R CrA region
using the Very Large Array.
The Class I protostar IRS 5b persistently showed polarized radio emission
with a constant helicity over 8 yr,
which suggests that its magnetosphere has a stable configuration.
There is a good correlation between the Stokes $I$ and Stokes $V$ fluxes,
and the fractional polarization is about 0.17.
During active phases the fractional polarization
is a weakly decreasing function of Stokes $I$ flux,
which suggests that IRS 5b is phenomenologically similar
to other types of flare stars such as RS CVn binaries.
The variability timescale of the polarized flux is about a month,
and the magnetosphere of IRS 5b must be very large in size.
The Class I protostar IRS 7A was detected once
in circularly polarized radio emission,
even though IRS 7A drives a thermal radio jet.
This detection implies
that the radio emission from the magnetosphere of a young protostar
can escape the absorption by the partially ionized wind
at least once in a while.
The properties of IRS 7A and IRS 5b suggests
that Class I protostars have
organized peristellar magnetic fields of a few kilogauss
and that the detectability of magnetospheric emission
may depend on the evolutionary status of protostar.
Also reported is the detection of circularly polarized radio emission
toward the variable radio source B5.
\end{abstract}

\keywords{ISM: individual (R Coronae Australis IRS 5) --- ISM: structure
          --- stars: flare --- stars: formation}

\section{INTRODUCTION}

Magnetic fields play an important role in star formation
(Andr{\'e} 1996; Feigelson et al. 2007; Bouvier et al. 2007).
This is especially true
in the protostellar accretion/evolution processes,
and peristellar magnetic fields are expected or even required
in many theoretical models:
Protostellar outflows are generated
through magnetocentrifugal ejection mechanism (e.g., Pudritz et al. 2007).
The magnetorotational instability may provide the viscosity
required to explain the dynamics of accretion disks (Balbus \& Hawley 1991).
Magnetospheres may force the protostellar spin
to be coupled with the rotation of circumstellar disk,
and the inner edge of accretion disks may be truncated
by the magnetosphere near the corotation radius
(Camenzind 1990; K{\"o}nigl 1991).
The transfer of material from the accretion disk to the protostar
may be done through magnetospheric accretion (Bouvier et al. 2007).
Despite all these expectations,
observational evidence of peristellar magnetic fields of protostars
is very difficult to come by.

Much of our knowledge about the peristellar magnetic fields
of young stellar objects comes from the observations of T Tauri stars,
especially radio continuum observations at centimeter wavelengths
(Andr{\'e} et al. 1992; Andr{\'e} 1996).
The key characteristics are the variability and circular polarization
of emission from nonthermal electrons.
The variability timescale is usually hours to days,
much longer than solar-type flares,
which implies that large-scale magnetic fields are involved.
Moderate degrees of circular polarization were observed,
suggesting that the emission mechanism is
gyrosynchrotron radiation from mildly relativistic electrons.
The detection of circular polarization at $\sim$5 GHz implies
a dipole-like large-scale magnetosphere
with a surface field strength of $\sim$1 kG (G{\"u}del 2002).

X-ray surveys of pre-main-sequence stars suggest
that the level of activity decays with stellar age
(Preibisch \& Feigelson 2005),
and we expect that Class I protostars may be
magnetically more active than T Tauri stars.
However, the free-free absorption by thermal wind
is supposed to obscure the magnetosphere of protostars in most cases
(Andr{\'e} et al. 1992; Andr{\'e} 1996).
The youngest object detected in circularly polarized radio emission
is IRS 5 in the R CrA region (Feigelson et al. 1998).
In fact, IRS 5 is the only known protostellar object
persistently displaying circularly polarized emission
(Forbrich et al. 2006; Miettinen et al. 2008).
Therefore, IRS 5 provides a wonderful opportunity
to study the magnetic activity of protostars.

To obtain high-quality images of the R CrA region,
we observed in the centimeter continuum with the Very Large Array (VLA)
of the National Radio Astronomy Observatory.
The main results were presented in Paper I (Choi et al. 2008).
In this paper, we present the polarimetry of the R CrA region.
We describe the data in \S~2.
In \S~3 we briefly report the results of the polarimetry.
In \S~4 we discuss the star forming activity of the radio sources
showing circularly polarized centimeter continuum emission.
A summary is given in \S~5.

\section{DATA}

We analyzed 14 data sets
to measure the circularly polarized flux densities
of the centimeter continuum sources in the R CrA region.
Details of the observations are described in Paper I.
See Table 1 of Paper I for the parameters of the VLA observing runs
and the observing track numbers.
(The track numbers are preceded by `Tr' throughout this paper.)
Maps of circularly polarized (Stokes $V$) intensity
were made using a CLEAN algorithm.
All the images presented in this paper
are corrected for the primary beam response,
and the lowest contour level roughly corresponds to the 3 $\sigma$ value,
where $\sigma$ is the rms of noise.

\begin{figure*}
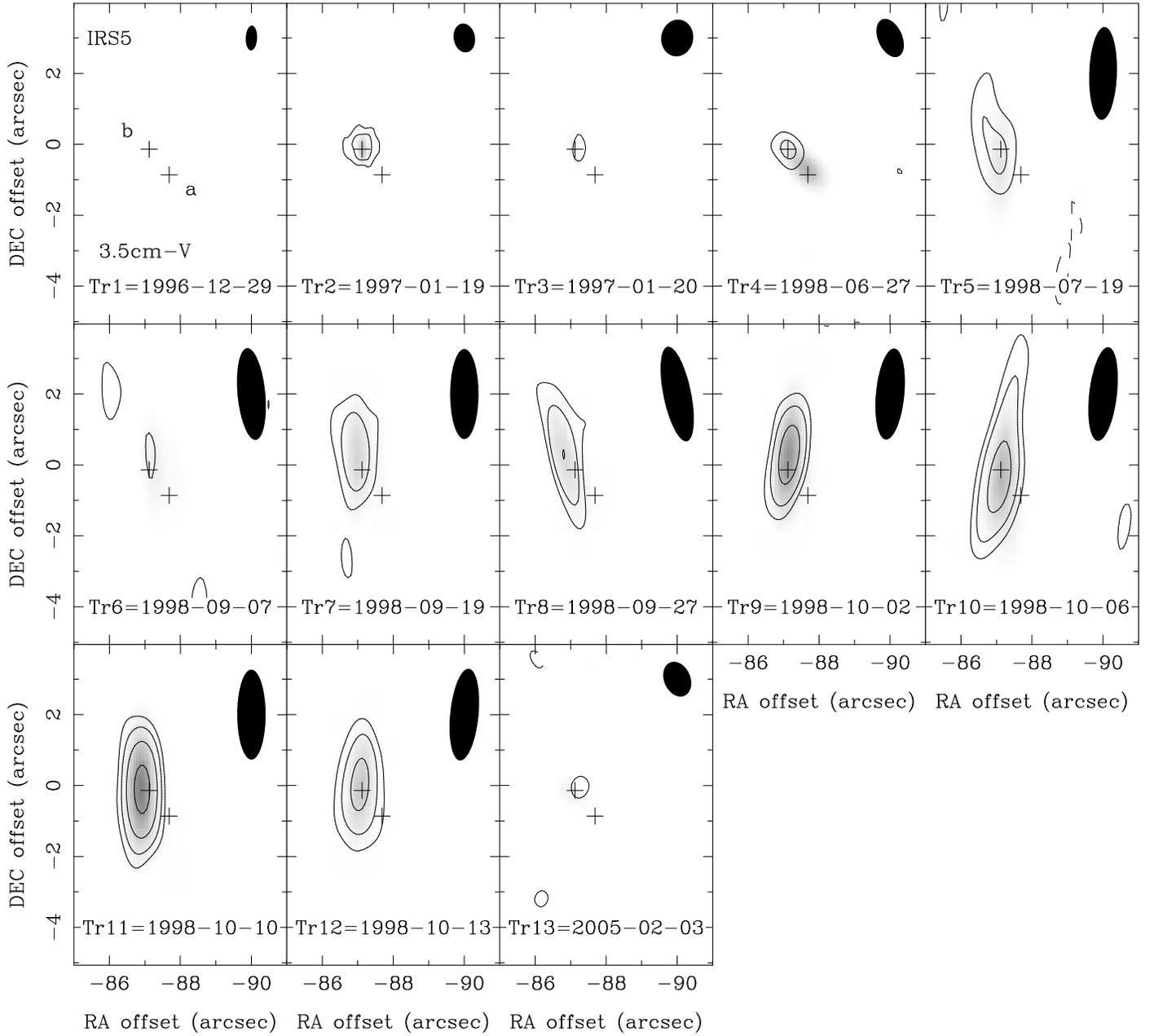

\ploteps{IRS5.XVnat.eps}{185mm}
\caption{\small\baselineskip=0.825\baselineskip
Stokes $V$ natural-weight maps ({\it contours})
of the 3.5 cm continuum toward IRS 5a/b.
Observation dates and track numbers are labeled.
A position correction was applied to the Tr 2 and 3 maps
(see \S~3.2 of Paper I).
The contour levels are 1, 2, 4, and 8 times 0.06 m\Jypb.
Shown in the top right corner are the synthesized beams
(see Table 1 of Paper I).
{\it Gray scale}:
Stokes $I$ maps as shown in Fig. 12 of Paper I.
{\it Plus signs}:
Peak positions of the 3.5 cm sources
in the uniform-weight Stokes $I$ map of Tr 1.}
\end{figure*}

The accuracy of polarization calibration of VLA data
is usually better than 0.5\% for compact sources.
For circular polarization measurements,
the beam squint can induce spurious signals,
which may affect sources located far away from the pointing center.
For snapshot images,
this effect can be as large as 20\% at the edge of the field of view.
For the maps made from the data of a whole track,
the circular polarization may be accurate to a few percent.
This issue will be discussed again at the end of \S~3.

\section{RESULTS}

Stokes $V$ flux was detected
toward three sources in the 3.5 cm maps
while none of the radio sources
showed detectable Stokes $V$ flux in the 6.2 cm maps.
The polarized emission of IRS 5 was reported previously
(Feigelson et al. 1998; Forbrich et al. 2006; Miettinen et al. 2008),
and we can now identify the source as IRS 5b.
Detections of polarized emission from IRS 7A and B5
are reported here for the first time.

\begin{figure}
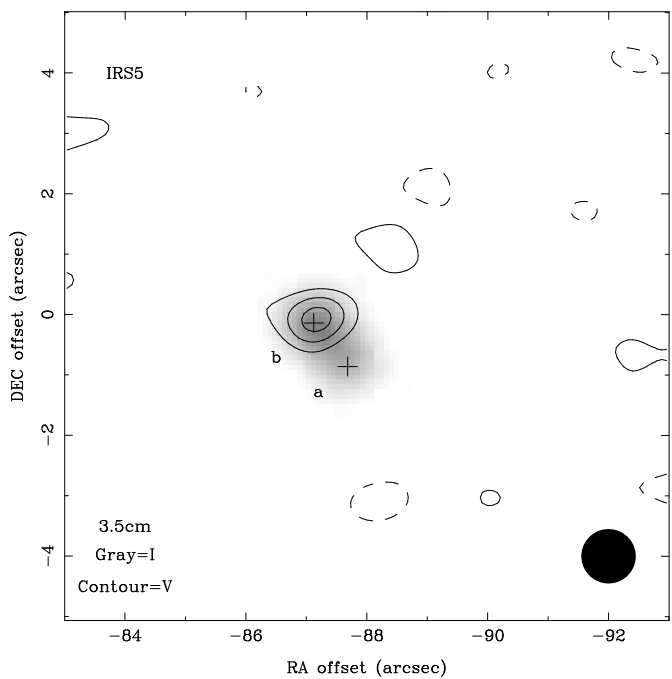

\ploteps{IRS5.XVavg.eps}{88mm}
\caption{\small\baselineskip=0.825\baselineskip
Stokes $V$ map ({\it contours}) of the 3.5 cm continuum
toward the IRS 5 region
from high-resolution data sets: Tr 2, 4, and 13.
A uniform-weight image was made from each data set.
Each image was corrected for the primary beam response
and convolved to an angular resolution of 0.9$''$
(shown in the bottom right corner),
and then the three images were averaged.
The contour levels are 1, 2, and 3 times 0.04 m\Jypb.
{\it Gray scale}:
Stokes $I$ map as shown in Fig. 10 of Paper I.
{\it Plus signs}:
Peak positions of the 3.5 cm sources
in the uniform-weight Stokes $I$ map of Tr 1.}
\end{figure}

\begin{deluxetable}{p{16mm}cccl@{~$\pm$~}l}
\tabletypesize{\small}
\tablecaption{Flux Densities of IRS 5b at 3.5 cm}%
\tablewidth{0pt}
\tablehead{
&& \colhead{$I$} & \colhead{$V$} \\
\colhead{Track} & \colhead{Date} & \colhead{(mJy)} & \colhead{(mJy)}
& \multicolumn{2}{c}{$\Pi_c$\tablenotemark{a}} }%
\startdata
Tr 1\dotfill  & 1996 Dec 29 & 0.25 $\pm$ 0.02 & \nodata & \multicolumn{2}{c}{$<$ 0.24\tablenotemark{b}} \\
Tr 2\dotfill  & 1997 Jan 19 & 1.82 $\pm$ 0.04 & 0.23 $\pm$ 0.02 & 0.126 & 0.011 \\
Tr 3\dotfill  & 1997 Jan 20 & 0.58 $\pm$ 0.04 & 0.09 $\pm$ 0.02 & 0.16  & 0.04  \\
Tr 4\dotfill  & 1998 Jun 27 & 0.53 $\pm$ 0.04 & 0.15 $\pm$ 0.02 & 0.28  & 0.04  \\
Tr 5\dotfill  & 1998 Jul 19 & 0.75 $\pm$ 0.04 & 0.16 $\pm$ 0.03 & 0.21  & 0.04  \\
Tr 6\dotfill  & 1998 Sep 07 & 0.45 $\pm$ 0.04 & 0.07 $\pm$ 0.03 & 0.16  & 0.07  \\
Tr 7\dotfill  & 1998 Sep 19 & 1.05 $\pm$ 0.04 & 0.22 $\pm$ 0.02 & 0.21  & 0.02  \\
Tr 8\dotfill  & 1998 Sep 27 & 0.91 $\pm$ 0.03 & 0.24 $\pm$ 0.02 & 0.26  & 0.02  \\
Tr 9\dotfill  & 1998 Oct 02 & 2.67 $\pm$ 0.04 & 0.34 $\pm$ 0.02 & 0.127 & 0.008 \\
Tr 10\dotfill & 1998 Oct 06 & 1.83 $\pm$ 0.04 & 0.42 $\pm$ 0.04 & 0.23  & 0.02  \\
Tr 11\dotfill & 1998 Oct 10 & 3.14 $\pm$ 0.05 & 0.60 $\pm$ 0.02 & 0.191 & 0.007 \\
Tr 12\dotfill & 1998 Oct 13 & 1.43 $\pm$ 0.04 & 0.31 $\pm$ 0.02 & 0.217 & 0.015 \\
Tr 13\dotfill & 2005 Feb 03 & 0.41 $\pm$ 0.03 & 0.08 $\pm$ 0.02 & 0.20  & 0.05  \\
\enddata\\
\tablecomments{Flux densities are corrected for the primary beam response.
               The Stokes $I$ fluxes are
               the same as given in Table 5 of Paper I.
               The Stokes $V$ fluxes are from the natural-weight maps.}%
\tablenotetext{a}{Fractional circular polarization.}%
\tablenotetext{b}{Stokes $V$ flux was undetected in Tr 1.
                  The upper limit of $\Pi_c$ was calculated
                  with the 3 $\sigma$ value of the $V$ map.}%
\end{deluxetable}

IRS 5b showed detectable polarization in most observing runs,
but IRS 5a was never detected in the Stokes $V$ flux maps (Figs. 1 and 2).
The Stokes $V$ flux densities from the archival data sets
are usually, but not always, consistent with the values
reported by Feigelson et al. (1998) and Forbrich et al. (2006).
Table 1 lists the total flux densities,
both Stokes $I$ and Stokes $V$ fluxes
(hereafter $I$ and $V$ fluxes, respectively),
and the fractional circular polarization, $\Pi_c = |V|/I$, of IRS 5b.

The $V$ flux from IRS 7A was detected
during the observations on 1998 July 19 (Tr 5).
The source was unresolved, and the peak position was
$\alpha_{2000}$ = 19$^{\rm h}$01$^{\rm m}$55.31$^{\rm s}$,
$\delta_{2000}$ = --36\arcdeg57$'$21.7$''$,
which agrees well (within 0.4$''$) with the position of IRS 7A
determined from a higher-resolution map (see Table 2 of Paper I).
The total flux was $V$ = 0.19 $\pm$ 0.02 mJy.
The fractional polarization cannot be exactly estimated
because IRS 7A has both compact and extended components
in the Stokes $I$ map.
Taking the intensity of the Stokes $I$ map with uniform weighting
as an upper limit on the $I$ flux from the central compact source,
the fractional polarization was $\Pi_c \gtrsim$ 0.07.

Circularly polarized emission from B5 was detected
during the observations on 2005 February 3 (Tr 13).
The source was unresolved, and the peak position was
$\alpha_{2000}$ = 19$^{\rm h}$01$^{\rm m}$43.28$^{\rm s}$,
$\delta_{2000}$ = --36\arcdeg59$'$12.3$''$,
which agrees well (within 0.1$''$)
with the peak position of the Stokes $I$ source from the same data set.
The total flux was $V$ = 0.26 $\pm$ 0.04 mJy,
and the fractional polarization was $\Pi_c$ = 0.12 $\pm$ 0.02.

The other radio sources in the R CrA region
were undetected in the $V$ flux.
These unpolarized sources are useful
in checking the quality of the polarization calibration.
If there had been a problem in the polarization calibration,
strong sources would show
a noticeable amount of apparent polarization systematically.
IRS 7B, IRS 1, and IRS 2 are suitable for this purpose.
Their $|V|$ intensity was always less than 0.05 m\Jypb,
and their $\Pi_c$ was less than 2 $\sigma$ level.
Therefore, we conclude that the polarization calibration was good
and that the detected $V$ flux densities of IRS 5b, IRS 7A, and B5
do not include any significant contribution from instrumental effects.

\section{DISCUSSION}

\subsection{IRS 5}

IRS 5 is an interesting example of protobinary system.
It is one of the youngest binaries with X-ray emission from both members,
indicating that both have high-energy magnetic activity
(Hamaguchi et al. 2008).
They displayed a concurrent enhancement of radio fluxes
and were suggested to be interacting (Paper I).
The projected separation of the binary is 0.9$''$,
which corresponds to 160 AU at a distance of 170 pc (Knude \& H{\o}g 1998).
IRS 5 as a whole is a Class I source
(Wilking et al. 1997; Nutter et al. 2005; Paper I),
but the accretion activity is weak.
Nisini et al. (2005) derived an accretion rate
of $\sim$3 $\times$ 10$^{-8}$ $M_\odot$ yr$^{-1}$ for IRS 5a
(the rate for IRS 5b may be even smaller),
which is much smaller than the accretion rate
of other Class I objects such as IRS 1 and IRS 2
(2 $\times$ 10$^{-6}$ $M_\odot$ yr$^{-1}$
and 3 $\times$ 10$^{-7}$ $M_\odot$ yr$^{-1}$, respectively).
IRS 5a/b may be Class I protostars
with the accretion activity halted temporarily.
(See White et al. 2007 for a discussion
on the accretion rate of Class I protostars in general.)
Nisini et al. (2005) suggested a spectral type of K5--K7V for IRS 5a,
and the spectral type of IRS 5b is probably later than that.

\subsubsection{IRS 5b}

While IRS 5b and 5a are comparable in the $I$ flux,
all the detectable $V$ flux can be attributed to IRS 5b (Figs. 1 and 2),
and separating them is important
in the analysis of the source properties.
Figure 3 shows the $I$ and $V$ light curves of IRS 5b,
and Figure 4 shows the relation between $I$ and $V$.
The flux uncertainty shown in Figure 3
is the statistical uncertainty listed in Table 1,
but one should also consider the uncertainty in the absolute flux scale,
usually about 1\%--2\% (VLA Calibrator Manual).%
\footnote{See http://www.vla.nrao.edu/astro/calib/manual.}
However, the flux scale cancels out
in the calculation of the ratio of flux densities,
and only the statistical uncertainty needs to be considered
for the fractional polarization.

\begin{figure}
\ploteps{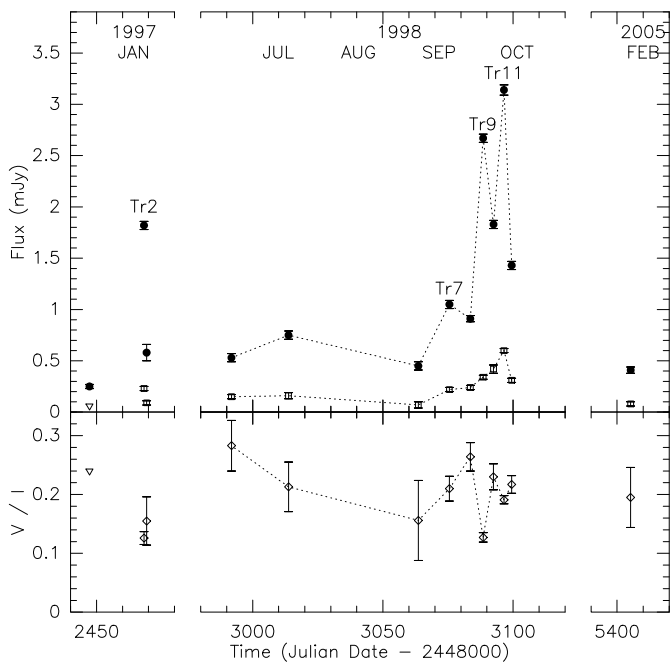}{88mm}
\caption{\small\baselineskip=0.825\baselineskip
Light curves of IRS 5b at 3.5 cm.
{\it Filled circles}:
Stokes $I$ flux densities.
{\it Open squares}:
Stokes $V$ flux densities.
{\it Diamond symbols}:
Fractional circular polarization.
{\it Open triangles}:
Upper limits (3 $\sigma$) at Tr 1.}
\end{figure}

\begin{figure}[t]
\ploteps{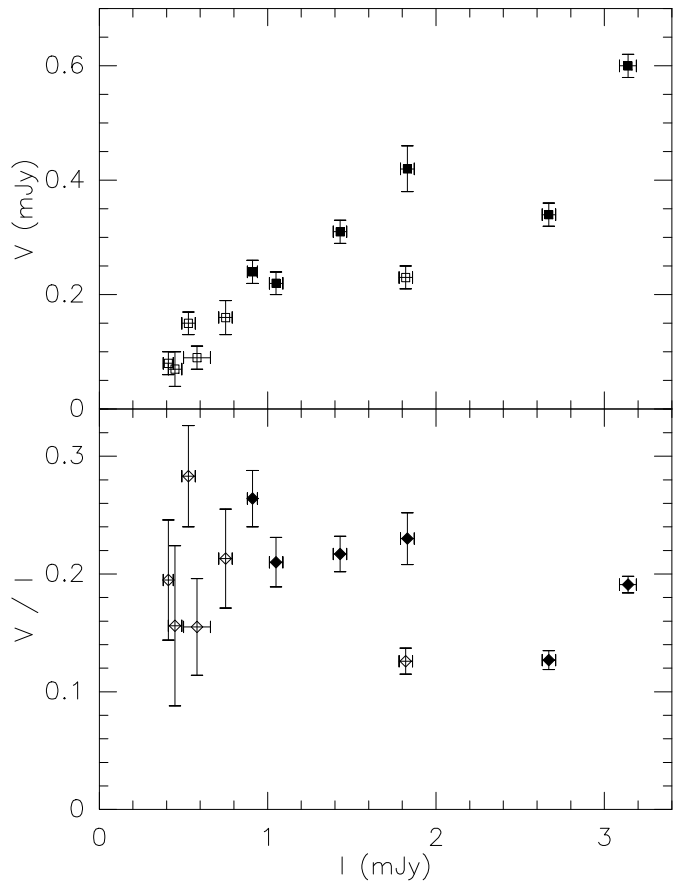}{88mm}
\caption{\small\baselineskip=0.825\baselineskip
Stokes $V$ flux ({\it top panel})
and fractional circular polarization ({\it bottom panel})
versus Stokes $I$ flux density of IRS 5b at 3.5 cm.
{\it Filled markers}:
Data points of the active phase in 1998 September--October.}
\end{figure}

The light curves of IRS 5b (Fig. 3) shows
that $V$ flux tends to be enhanced during the flare events.
Such a trend is not obvious if IRS 5a/b are unresolved:
Forbrich et al. (2006) thought
that the $V$ flux appeared uncorrelated to the $I$ flux
because they were distracted by the flux variability of IRS 5a.
Miettinen et al. (2008) also reported
that the fractional polarization can be as high as 0.33,
but their angular resolution was not high enough
to make a clear interpretation.
By separating IRS 5b from 5a,
it is possible to make a detailed interpretation of the polarimetry results,
as described below.

\subsubsubsection{Helicity}

The $V$ flux was always positive (right-handed)
during the 8 yr of monitoring.
The helicity of emergent radiation is related
to the configuration of magnetic fields
(Morris et al. 1990).
The constant helicity suggests
that the magnetic field geometry responsible for the polarization
is very stable and has a large scale (Mutel et al. 1987).
Here, the large-scale geometry means
a star-sized global magnetosphere or an even larger structure.
VLBI observations revealed
that some RS CVn binaries have circumbinary halo
and that some T Tauri stars have radio structures
as large as 15 times the stellar radius
(Mutel et al. 1985; Phillips et al. 1991).
Dipole-like magnetospheric models are often employed
to explain the radio observations of RS CVn binaries, T Tauri stars,
and other radio stars (G{\"u}del 2002).
In the case of IRS 5b,
the structure of the magnetosphere can be more complicated
because it may possess a circumstellar disk.

\subsubsubsection{The Active Phase in 1998 September--October}

As mentioned above, 
there is an overall correlation between the $I$ and $V$ light curves.
During the active phase in 1998 (Tr 7--12),
the signal-to-noise ratio is high enough
that a more detailed description of the light curves is possible.
While the $I$ curve shows a relatively rapid fluctuation,
the $V$ curve varies slowly and shows only one peak (Fig. 3).
The difference in the fluctuation timescale suggests
that the $I$ flux may respond to the energy injection relatively quickly
and the $V$ flux may come from a population of electrons
having a longer relaxation timescale.
This behavior can be explained by a core-halo model
(Mutel et al. 1985; see the discussion below).

\subsubsubsection{$I$-$V$ Correlation}

When only IRS 5b is considered,
the correlation between $I$ and $V$ fluxes is quite good (Fig. 4).
The $V$ flux is roughly proportional to the $I$ flux.
The average value of $\Pi_c$ is 0.17, and the standard deviation is 0.04.
This good correlation strongly implies
that the $I$ and $V$ fluxes come from the same source of energy
and that the radiation mechanisms in the active phase
and in the quiescent phase are basically the same.

If the quiescent radio emission is caused
by frequent injections of energetic electrons
through microflares (G{\"u}del 2002),
radio luminosity is expected to be related to X-ray luminosity.
The 3.5 cm flux density of IRS 5b in the quiescent phase
is $\sim$0.5 mJy (Paper I),
and the radio luminosity is
$L_R \approx$ 1.7 $\times$ 10$^{16}$ erg s$^{-1}$ Hz$^{-1}$.
The X-ray luminosity of IRS 5b is
$L_X$ = 1.9 $\times$ 10$^{30}$ erg s$^{-1}$ (Hamaguchi et al. 2008).
Then the ratio is $L_X/L_R$ = 1.1 $\times$ 10$^{14}$ Hz,
which is consistent with the empirical value of 10$^{15\pm1}$ Hz
for active stars (G{\"u}del 2002).
(See also the $L_R$-$L_X$ relationship
discussed by Feigelson et al. 1998 and Forbrich et al. 2006.)
Therefore, the quiescent radio emission of IRS 5b
is phenomenologically similar to what is observed in active late-type stars.

Figure 4 shows
that the relation between $\Pi_c$ and $I$ is unclear
when all data are considered.
If the attention is restricted to the active phase only ($I >$ 0.8 mJy),
the scatter of $\Pi_c$ is smaller,
and $\Pi_c$ seems to be a weakly decreasing function of $I$.
This trend is clearer if only the active phase of 1998 is considered.
Such a behavior was also found in some radio active stars:
the fractional circular polarization of RS CVn binaries and Algol
decreases with increasing luminosity (Mutel et al. 1987, 1998).
This anticorrelation was explained using a model
in which the quiescent emission comes
from an optically thin region (halo)
of low Lorentz factor ($\gamma \lesssim$ 5)
while the intense flare emission comes from an optically thick region (core)
of more energetic electrons (Mutel et al. 1985, 1987).

\subsubsubsection{Strength of Magnetic Fields}

The emission mechanism of circularly polarized radio emission
at centimeter wavelength
is nonthermal gyrosynchrotron emission,
and the detection of $V$ flux at centimeter wavelength implies
that the strength of the magnetic fields in the source region
is a few hundred gauss or larger (G{\"u}del 2002).
The turnover frequency, $\nu_{\rm peak}$,
is a good indicator of the field strength.
For example, White et al. (1989) derived a simple relation
for a dipole source with typical parameters of M dwarf stars,
\begin{equation}
   \nu_{\rm peak} = 4.2 (B_0/1000)^{0.76}\ {\rm GHz},
\end{equation} 
where $B_0$ is the surface magnetic field strength in G.
In the quiescent phase, IRS 5b is brighter than IRS 5a,
and the spectral index of IRS 5b in the 4.9--8.5 GHz range
is flat or slightly negative (Paper I).
Then the turnover frequency of IRS 5b
may be roughly in this frequency range,
and the inferred field strength is $B_0 \approx$ 2 kG.

\subsubsection{IRS 5a}

While IRS 5a is more active in X-rays than IRS 5b,
as it frequently showed solar type X-ray flares (Hamaguchi et al. 2008),
it never showed polarized radio emission.
The nondetection of circular polarization of IRS 5a
is not simply owing to the relatively low $I$ flux,
because $V$ flux was undetected even in the strong flare event of Tr 4.
The 3 $\sigma$ upper limit on the polarization fraction is 0.03.
Therefore, while both radio polarization and X-ray emission
are related to magnetic activity,
there is no simple correlation
between polarized radio flux and X-ray flares.

IRS 5a and 5b are quite similar in many aspects.
They have comparable X-ray luminosities and almost identical X-ray spectra
in the quiescent phase (Hamaguchi et al. 2008).
Since they are in a binary system,
they must have a similar age and are in the same environment.
Then what causes the difference in the polarization of radio emission?
A possible answer is the source geometry.
It is well known that the fractional circular polarization of RS CVn binaries
is dependent on the inclination angle of the system
(Mutel et al. 1987, 1998; Morris et al. 1990).
If this dependency is applicable to the IRS 5 system, it suggests
that the magnetosphere of IRS 5b may have a pole-on configuration
while that of IRS 5a may be equator-on.

\subsubsection{Peculiarity of IRS 5}

IRS 5b is the only known protostellar object
persistently displaying circularly polarized radio emission.
While this gives us a good opportunity
to study the magnetic activity of protostars,
it raises a question: why IRS 5b is an exception to the rule
that the protostellar magnetosphere
is hidden behind the free-free photosphere
(the surface where the optical depth of free-free radiation equals unity)?
That is, IRS 5b has either an unusually small free-free photosphere
or an unusually large magnetosphere.
Possible answers may include
a small mass-loss rate (decreased ionized wind),
a circumstellar disk with an inner hole (lack of disk-driven warm wind),
and/or large-scale magnetic fields generated by disk dynamo
(a disk-scale magnetosphere)
(Andr{\'e} et al. 1992, Tout \& Pringle 1996).

Another unusual trait of IRS 5 is
that it is the only known protostellar binary system
with both members detected in X-rays (Hamaguchi et al. 2008).
An interesting issue is the effect on the ionization of disk material
because an accretion disk in such a system
is illuminated by two X-ray sources.
In terms of energy, only a small fraction of the X-ray photons from a star
can illuminate the disk around its companion.
(For example, a face-on disk with a radius about a third of binary separation
would intercept less that 3\% of the luminosity of the companion star.)
However, this extra X-ray source can irradiate the disk from above
if the binary system is not coplanar,
and the ionizing photons can penetrate into the disk midplane
relatively easily.
In addition, such an effect can generate some asymmetry in the disk.
Study of the dynamics and chemistry of accretion disks
in such an environment would be useful
because many stars form as members of multiple systems.
A circumstellar disk in a multiple system
may go through the X-ray irradiation
not only from the host star but also from its companion
at some point during the protostellar evolution.

\subsection{IRS 7A}

The evolutionary status of IRS 7A is unclear (see \S~5.1.1 of Paper I)
because the spectral energy distribution (SED) is poorly constrained.
IRS 7A is definitely not a Class II object (T Tauri star).
The X-ray characteristics of IRS 7A is similar to that of Class I objects
(Hamaguchi et al. 2005).
The mid-IR SED of IRS 7A is similar
to that of the Class 0/I transitional object IRS 7B
(also see the color-color diagram in Paper I),
but IRS 7A is less embedded than the nearby Class 0 source SMA 2
(Groppi et al. 2007).
IRS 7A displays an outflow activity (Paper I),
suggesting that accretion is also active.
Considering all these facts,
IRS 7A is most likely a protostar in the early part of the Class I stage.
Then it is the youngest known object
detected in circularly polarized radio emission,
or a protostar showing an evidence of magnetosphere
with an organized field configuration
and a field strength of the order of 10--100 G.

\begin{figure}
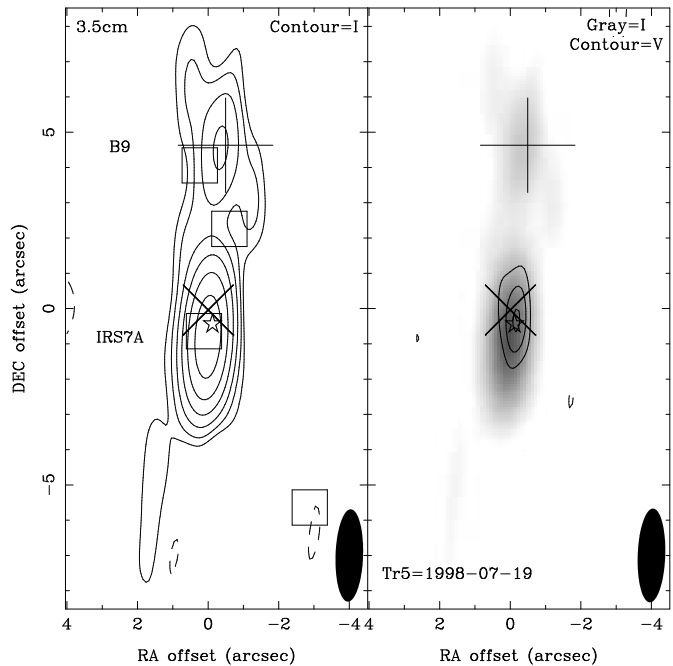

\ploteps{IRS7A.XV.eps}{88mm}
\caption{\small\baselineskip=0.825\baselineskip
Natural-weight maps of the 3.5 cm continuum
toward the IRS 7A region from track Tr 5.
Shown in the bottom right corner is the synthesized beam:
FWHM = 2.6$''$ $\times$ 0.8$''$ and P.A. = --2\arcdeg.
{\it Left panel}:
Stokes $I$ map.
The contour levels are 1, 2, 4, 8, 16, and 32 times 0.06 m\Jypb.
{\it Right panel}:
Stokes $V$ map ({\it contours})
superposed on the Stokes $I$ map ({\it gray scale}).
The contour levels are 1, 2, and 3 times 0.06 m\Jypb.
{\it Plus sign}:
The 1.1 mm continuum source SMA 2 (Groppi et al. 2007).
{\it Cross}:
X-ray source CXOU J190155.3--365722 (Hamaguchi et al. 2005).
{\it Star symbol}:
Mid-IR source IRAC 5 (Groppi et al. 2007).
{\it Squares}:
The 7 mm continuum sources (Choi \& Tatematsu 2004).}
\end{figure}

Circularly polarized flux from IRS 7A was detected
only once out of the 13 observing tracks reported here (Fig. 5).
In addition, Suters et al. (1996) and Forbrich et al. (2007)
also made Stokes $V$ maps of the R CrA region
but did not report any detection at the position of IRS 7A.
Therefore, IRS 7A must be displaying
a detectable circular polarization only rarely.
However, at the epoch of the detection (Tr 5),
the $I$ flux was not particularly high.
In fact, the $I$ flux was on the low side of the scatter
(see Table 4 of Forbrich et al. 2006).
Even if we consider the intensity
at the position of IRS 7A in uniform-weight maps
(to minimize the blending of flux from the extended outflow),
the $I$ intensity at Tr 5 is on the weaker side.
This difference between $I$ and $V$ fluxes obviously shows
that most of the $I$ flux comes from a thermal source (Paper I)
and only a fraction of $I$ flux can be directly associated with the $V$ flux.
Therefore, if the angular resolution had been high enough
to separate the thermal and nonthermal sources,
the fractional circular polarization of the nonthermal emission
must have been much higher than 0.07 (see \S~3).

In contrast to the radio $V$ flux,
the X-ray emission of IRS 7A was detectable most of the time
(Hamaguchi et al. 2005; Forbrich et al. 2006, 2007).
Then an active magnetosphere probably exists all the time,
and nonthermal radio emission may persist at some level.
If so, why is the circular polarization usually undetectable?
In fact, the nondetection is not surprising
because the radio emission from the peristellar region of protostars
is supposed to be obscured owing to the free-free absorption
by the partially-ionized thermal winds (Andr{\'e} et al. 1992).
Then the question becomes how the radiation from the magnetosphere
survived the absorption at the epoch of detection?
Several possibilities can be suggested.
First, the magnetosphere momentarily grew very large.
Either the whole magnetosphere expanded
or a portion of it stretched out.
Second, the free-free photosphere shrank temporarily.
This rarefication of wind can happen
if the mass-loss rate (hence the accretion rate) decreases for some reason.
Third, the thermal wind was far from spherical
and allowed an occasional direct view of the magnetosphere.
Such a favorable line-of-sight can be achieved
if the wind is collimated and changes directions.
In principle, these possibilities can be tested using the VLBI technique
although the transient nature of the phenomenon will be a major difficulty.

\subsection{Magnetic Fields of Protostars}

If we take IRS 7A and IRS 5b as indicative examples
of the evolution of magnetically active Class I objects,
for lack of other examples, a plausible picture emerges:
Most X-ray emitting Class I protostars may have magnetospheres.
Younger ones, such as IRS 7A, may hide
the existence of organized magnetic fields most of the time
because the mass-loss rate is usually high
and the wind is optically thick.
Older protostars, such as IRS 5b, have a decreased level of mass loss,
the wind becomes optically thin,
and the magnetosphere becomes easily visible.
(See the discussion by Andr{\'e} et al. 1992.)
The obvious next question is
whether such magnetospheres exist around Class 0 protostars.
Since there is no bona fide Class 0 object with detectable X-ray emission,
the answer to this question may be hard to get.
(This issue may be somewhat controversial
because IRS 7B is around the verge of the Class 0 protostar category.
See Hamaguchi et al. 2005
for more discussions on the implications of X-ray emission from IRS 7B.)
A good starting point may be monitoring observations
of X-ray-emitting young protostars such as IRS 7B.
Another possible strategy is monitoring observations
of Class 0 protostars that are weak in free-free emission.
In this respect, the detection of circularly polarized 6 cm emission
from the very low luminosity object L1014-IRS
warrants further investigations of similar objects (Shirley et al. 2007).

The protostellar magnetosphere plays an important role
in the circumstellar accretion
and the accretion of matter in the inner disk.
Theoretical models and observational characterization
are relatively well developed for T Tauri stars,
but not much is known about protostellar magnetospheres.
Results of radio polarimetry, as reported in this paper,
are beginning to provide valuable information on this important issue.
The radio properties of IRS 7A and IRS 5b imply
that the magnetic activity of Class I protostars
are qualitatively similar to those of T Tauri stars.
Future telescopes with better sensitivity and higher angular resolution
should be able to help us understand
the role of magnetic processes in protostellar evolution,
and monitoring observations are crucial
because of the time variability of these sources.

\subsection{Frequency Dependence of Polarization}

All the three sources displaying circularly polarized emission at 3.5 cm
did not show any detectable polarization at 6.2 cm.
Such a frequency dependence is consistent with the observed properties
of other magnetically active stars such as RS CVn systems:
the fractional polarization tends to increase with frequency
in the 5--15 GHz range
(White \& Franciosini 1995; Garc{\'\i}a-S{\'a}nchez et al. 2003).
However, it is not easy to explain this behavior
using simple theoretical models of gyrosynchrotron radiation,
and inhomogeneous models were suggested
(Jones et al. 1994; White \& Franciosini 1995).
It is unclear how magnetically active protostars would behave
over a wide range of frequency,
and multifrequency observations of IRS 5b will be interesting.

\subsection{B5}

B5 is a radio source with rapid flux variability
(Brown 1987; Suters et al. 1996).
It is not an extragalactic source,
but its nature is poorly known (Paper I).
Feigelson et al. (1998) suggested that B5 is a brown dwarf,
but we cannot rule out the possibility
that it may be a background object
located much further away than the Cr A cloud.

\begin{figure}
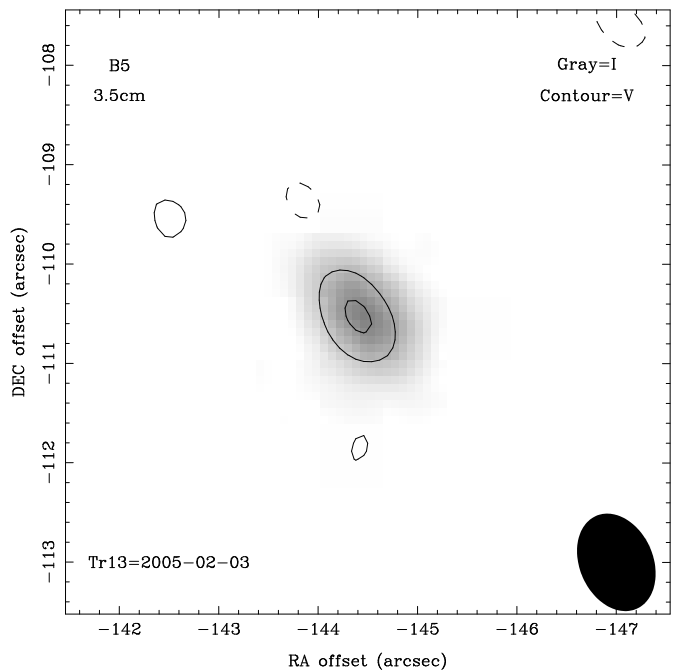

\ploteps{B5.XV.eps}{88mm}
\caption{\small\baselineskip=0.825\baselineskip
Stokes $V$ natural-weight map ({\it contour}) of the 3.5 cm continuum
toward the B5 region from track Tr 13.
The contour levels are 1 and 2 times 0.12 m\Jypb.
Shown in the bottom right corner is the synthesized beam:
FWHM = 1.0$''$ $\times$ 0.7$''$ and P.A. = 23\arcdeg.
{\it Gray scale}:
Stokes $I$ map as shown in Fig. 14 of Paper I.}
\end{figure}

Circularly polarized emission from B5 was never reported before
and detected only once (Tr 13) out of the 13 observing tracks (Fig. 6).
The $I$ flux during Tr 13 was larger than the flux at the other epochs
by a factor of $\sim$3 (see Fig. 2 of Paper I).
The fractional polarization during Tr 13
was also higher than the other epochs
because polarized flux must have been detectable most of the time
if $\Pi_c$ were constant.
This enhancement of $I$ and $\Pi_c$ during Tr 13 suggests
that B5 seems to have gone through an outburst
of highly polarized radio emission.
Interestingly, the quiescent emission is also nonthermal
because the spectral index during Tr 1 is highly negative
(see Table 3 of Paper I).
Then the radio emission structure of B5 must be peculiar:
the quiescent emission may be
synchrotron radiation from a large-scale structure,
and the intense flare emission may be gyrosynchrotron radiation
from a compact structure, probably a peristellar magnetosphere.
Since the 6.2 cm source is extended
and has a position offset with respect to the 3.5 cm source (Paper I),
the large-scale structure is probably an outflow.

If B5 is a brown dwarf as suggested by Feigelson et al. (1998),
it joins the small list of substellar objects
detected in circularly polarized radio emission
(Berger et al. 2001; Berger 2006).
These objects show much higher radio luminosity
than what is expected from the X-ray luminosity,
which may explain the nondetection of B5 in X-rays.
It is not well understood
how such cool dwarfs can generate large-scale magnetic fields
(Berger 2006).

On the other hand, the peculiar emission structure
suggests a quite different possibility:
B5 may be an exotic object
containing a magnetically active compact source
that is ejecting material at a highly relativistic speed.
But it is certainly not an X-ray binary
because no X-ray detection was ever reported.
The nature of B5 is still an open question.

\section{SUMMARY}

The R CrA region was observed using the VLA in the 3.5 and 6.2 cm continuum
to image the young stellar objects
and study the star forming activity in this region.
In addition, archival VLA data from recent observations
made in extended configurations were analyzed.
The results based on the imaging of Stokes $I$ intensity
were reported in Paper I,
and the results of Stokes $V$ imaging are reported in this paper.
A total of 3 sources was detected in the 3.5 cm Stokes $V$ flux
while none was detected in the 6.2 cm circularly polarized continuum.
The main results are summarized as follows:

1.
Circularly polarized emission from IRS 5 was detected most of the time,
and the exact source of the $V$ flux was identified as IRS 5b.
When IRS 5a and 5b are separated,
there is a good correlation between the $I$ and $V$ fluxes of IRS 5b.
The circular polarization is present at a $\sim$17\% level,
even in the quiescent phase,
as long as the sensitivity is high enough to detect the $V$ flux.

2.
The $V$ flux from IRS 5b was always positive
over the 8 yr of monitoring,
suggesting that the magnetosphere has a stable configuration.
During the active (flaring) phases,
when the signal-to-noise ratio is high enough,
the fractional polarization is a weakly decreasing function of $I$ flux.
The active phases of IRS 5b last about a month or longer,
and the light curve of $V$ flux varies more slowly than that of $I$ flux,
which suggests
that the source of the circularly polarized emission is huge in size.
These properties suggests
that IRS 5b is phenomenologically similar to other types of flare stars
and has a very large magnetic structure.

3.
IRS 5a was never detected in the circularly polarized radio continuum
even during flares.
The cause of the difference between IRS 5a and 5b is unclear,
but a geometrical effect is possible.

4.
Circularly polarized flux from IRS 7A was detected once
out of the 13 observing tracks analyzed.
The fractional circular polarization was at least 7\%.
Since IRS 7A drives a thermal radio jet,
it is unclear how the polarized flux
was not completely absorbed by the thermal wind.

5.
Considering the properties of IRS 7A and IRS 5b, we suggest
that most (X-ray emitting) Class I protostars may have magnetospheres
and that the detection of radio emission from the magnetosphere
becomes easier as the protostar evolves and the mass-loss rate decreases.
Future surveys of similar Class I sources
with the Expanded Very Large Array
will be very exciting
because the continuum sensitivity will be improved by an order of magnitude,
and it may be possible to test the hypothesis above.

6.
Circularly polarized emission from B5 was detected once,
and the fractional circular polarization was 12\%.
B5 was in an outburst when the $V$ flux was detected.
While the detection of circular polarization shows
that there is an organized magnetic field structure,
the nature of B5 is still unclear.

\enlargethispage{-15\baselineskip}

\acknowledgements

We thank Y. L. Shirley and J. Cho for helpful discussions and suggestions.
M. C. was supported by the Korea Science and Engineering Foundation (KOSEF)
through the grant of the basic research program R01-2007-000-20196-0.
K. H. is supported by the NASA Astrobiology Program under CAN 03-OSS-02.
J.-E. L. gratefully acknowledges the support by KOSEF
under a cooperative agreement with the Astrophysical Research Center
for the Structure and Evolution of the Cosmos (ARCSEC).
The National Radio Astronomy Observatory is
a facility of the National Science Foundation
operated under cooperative agreement by Associated Universities, Inc.


\end{document}